# A human centered perspective of E-maintenance

Allan Oliveira, Regina Araujo
Computer Department
Federal University of São Carlos
São Carlos, São Paulo, Brazil
allan_oliveira@dc.ufscar.br, regina@dc.ufscar.br

**Abstract**

*E-maintenance is a technology aiming to organize and structure the ICT during the whole life cycle of the product, to develop a maintenance support system that is effective and efficient. A current challenge of E-maintenance is the development of generic visualization solutions for users responsible for the maintenance. AR can be a potential technology for E-maintenance visualization, since it can bring knowledge to the real physical world, to assist the technician perform his/her work without the need to interrupt to consult manuals for information. This paper proposes a methodology for the development of advanced interfaces for human aware E-maintenance so that complex maintenance processes can be made safer, better quality, faster, anytime and anywhere.*

## 1. Introduction

Over the years the amount of data available in the industry has grown to a state that ICT (Information and Communication Technologies) has become vital for the industry. Now, with all the data in digital form, and a large part also online (accessible through the internet), it is possible to create information from these data, and knowledge from the information. From knowledge many types of systems can be developed to assist industry to manage and improve their services and products. E-manufacturing, E-business and E-maintenance are emerging solutions to support industry.

Producers of complex products (such as airplanes, hydroelectric power plant) have an increasing demand from customers to improve and increase dependability, availability, safety, sustainability, cost-effectiveness, operational flexibility, decrease Life Support Cost (LSC) and provide worldwide support at anytime [6]. At the same time the costs for such features are increasing since data, demand, and quality is also increasing.

Currently the maintenance system is a support system which has the function to provide data about the maintenance of the product. Although the ICT has improved and online information can be used during the maintenance process, these information need to be integrated to form a powerful maintenance system with all the information about the product, like technical information (product data and information), maintenance plans, fault diagnosis, real time inspection, etc. The integration of these data in a system is the objective of the E-maintenance.

There are many definitions for E-maintenance, one of the classics being a technology that matches ICT with the necessity to keep machines running properly, efficiently and safely [14], dealing with the ever increasing information flow and complexity of actual technical systems [6]. It organizes and structures the ICT during the whole life cycle of the product, to develop a support system that is effective and efficient.

Advantages of E-maintenance include [14]:

• New maintenance types and solutions: online (remote) cooperative maintenance makes it possible to quickly learn and share solutions with experts around the globe.

• Improved maintenance tools and support: maintenance documentation and online support can be created, so that adjustments are shared.

• Enriched maintenance activities: "E-diagnosis" enables experts to perform online fault diagnosis, and since it is possible to keep contact with the manufacturers of a product, the repair time is reduced.

In contrast these same advantages generate several challenges to be overcome. First it is necessary to restructure the human resources to pace with the new information flow and overall structure. This is also part of the "total productive maintenance" holistic planning process.

Then several cross-platform information integrations have to be developed, such as data transformation mechanism, communication protocols and safe network connections. Since data is expect to come from several different sources, such as ICT systems, Internet of Things (IoT), Web and Sensors, a powerful mechanism for data analytics, data fusion and interpretation is necessary.

Another challenge is the development of generic representation solutions for the development of systems. Furthermore, generic visualization solutions for the engineer or technical responsible for the maintenance and for structuring the maintenance data are required.

One visualization solution that is being used in the maintenance field is Augmented Reality. Augmented Reality (AR) is the research area in which synthetic computer-generated objects supplement or compose real environments, providing users with a more resourceful tool to visualize their surroundings and related information without space or mobility constraints [15].

AR can be a potential technology for E-maintenance visualization, since it can bring knowledge to the real physical world, to assist the technician perform his/her work without the need to interrupt to consult manuals for information. Several solutions for Augmented Reality in maintenance have emerged, some of them are: ARVIKA [10][24], AMRA [7], STARMATE [20], ULTRA [18], BMW [17], T.A.C. [5], ARIMA [2] and ARMAR [11].

Using AR interface in the maintenance process, users can become aware of their surroundings and tasks in real time. This means that not only this interface can enhance and facilitate the users´ job, but it has also the potential to ensure users´ and machines´ safety, showing the possible dangers of executing current operations. It is also possible to ensure several levels of user experience depending on users´ expertise on the subject and profile. For instance, an engineer or a technician could see different things in the same AR application, since they have different levels of expertise, and the system can record and adapt to each individual users´ previous experiences.

For this scenario to be possible, it is important to understand the process of maintenance, considering both laboratory and field study, so that models, methodologies, techniques and tools can be developed.

Therefore the presented work proposes to create a methodology for the development of enriched and valuable interfaces to E-maintenance that can assist technicians and engineers (hereby known as just "users") in their tasks of complex maintenance anytime, anywhere.

The methodology will be used to create an AR application maintenance model for each equipment available, therefore it has to be fast to use, have reusable components and able to create complete AR applications.

## 2. Related Work

There are several works, some already cited before, for creation and visualization of AR focused in maintenance and other procedural tasks in industrial processes. Most tend to focus on building an application in a real case scenario and proving their validity. Some focus on the user experience, and others in creating new modes for interaction or new paradigms for the maintenance task.

One of the first works to propose the use of an Augmented Reality interface for equipment maintenance is from Lipson et al. [13]. They created the Online Guided Maintenance approach, which uses a web link to obtain the equipment maintenance content, and feed the AR application with it.

T.A.C. (Tele-Assistance-Collaborative) [5] aims to combine remote collaboration and industrial maintenance, creating systems which implement co presence collaboration between the operator executing the maintenance and an assisting co worker (normally of a higher level of understanding of the equipment). T.A.C. invented a new interaction paradigm, P.O.A. (Picking Outlining Adding), which is based in three points: the ability to quickly point to an object in the 3D scene using a mouse, the ability to highlight an object outline, and the ability to produce an animation to demonstrate the execution of a task.

Arvika [10][24] is perhaps the most advanced (or applied) work for AR in maintenance and industrial contexts to date. The project had a vision of providing user centric mobile and stationary AR solutions for industrial applications. Its architecture had three main focuses: tracking (and visualizing AR), information provision from IT systems for the application, and user interaction.

Like Arvika, some works already focus on using the IT structure already existent in the industry to improve their AR solutions. ARIMA [2] is platform for AR maintenance based on collaboration between technicians and experts. The work is similar to T.A.C. but with a greater focus in integrating AR in an existing E-maintenance system.

The work of Espindola et al. [8] proposes a methodology for the visualization of OSA-CBM architecture layers during maintenance. OSA-CBM is an open architecture proposed by MIMOSA for the implementation of condition-based maintenance systems. They proposed a module for OSA-CBM to visualize data using mixed reality (2D and 3D interfaces).

Lastly there is an interesting work from BMW [22] that demonstrates a view of how in the future technical documentation will be prepared for Augmented Reality using metadata based authoring. The core idea is to adapt the already necessary creation of maintenance documentation to virtual or augmented reality compatible format, using the metadata of the process (dis-/assembly), the tools involved, the part and its geometry, and connected peripherals.

## 3. Project

Although AR development solutions evolved a lot, going from simple libraries like ARToolkit [12], to components frameworks as Studierstube [19], and now commercial solutions as Metaio SDK, Layar SDK and Qualcomm SDK, it is still considered an "experts only" solution. Part of this is because few authoring tools were developed, and each focusing on their own domain/problem.

Regarding the domain, an AR solution that aims to be a part of the industrial workflow must be integrated with the E-maintenance system and provide an intelligent AR-oriented information system [10].

An AR-maintenance application will have to understand the data provided by the E-maintenance system, and apply it to create an interface in the real world that enhances the user ability and experience in the maintenance task. Also this application has to serve as a visualization interface for all the data produced by the E-maintenance system, for technicians and engineers to accomplish their task in a fast and easy way.

Therefore, the main objective is to create and evaluate a methodology for the development of advanced interfaces for human aware E-maintenance so that complex maintenance processes can be made safer, better quality, faster, anytime and anywhere.

Some of the questions the work intends to answer include:

About the user:

What are the information technicians and engineers use for maintenance? What particular sources are used, and when and how are they used in the maintenance process? What are the most and least used information? How is the task executed? Does the environment influence the execution of the task? How much and how does the expertise level of the user influence his rate of consultation from multiple sources and how much is the time spent in each consultation? How long does it take to execute each task related to the user expertise? How efficient is AR remote collaboration? How much efficient and usable an AR interface can be? Does concentrating all information in one system helps being more efficient? Is the training using Augmented and Virtual Reality satisfactory? Does the interface enhance your awareness? What about Team awareness?

About the interface:

What is the current interface used to show data to users? What is the efficiency and usability of this interface? How does it help the user? What type of graphical information is used in this interface (graphics, text, pictures, video, sound)? What does the user think about each type of information, what are the most useful and easiest to learn and understand? What are the most complete and used ones? How does this interface adapt to the expertise level of the user? How is the data fusion displayed in the interface? Is the component-based interface enough to display all necessary and wanted information? Is model based design sufficient to create AR interfaces?

About the data:

What are the sources for the data used in the maintenance process? What are the types of data used in the interface? How are these data obtained? What is the structure of these data in the source? How are these data structured in the interface? Is it necessary to merge part of the data for the interface? What are the most useful data sources to the user? What are the best fusion techniques to apply to the data?

## 4. Proposed Methodology

The methodology is divided in 6 phases:
1. Gather and structure data
2. Creating a behavior and task model
3. Identifying contexts for situation awareness and team situation awareness
4. Designing user models
5. Modeling the user interface

The final result is an AR application maintenance model for each equipment, which is a specification of the AR interface describing the maintenance process of the equipment and the interface displayed.

### A. Gather and structure data

This phase is when all data about the equipment, and its maintenance process, is centralized.

Currently all the data (or at least most of it) for the creation of intelligent AR applications for maintenance already exist in the industries, but it is not used for this purpose.

One of the goals of E-maintenance is to structure these data so that it can be available for users as they request. The main data for the development of an AR system is:

- Description of the equipment to receive maintenance: basic data about the equipment, maintenance documentation manuals, all the pieces and tools that are used to assembly it, 3d models, pictures from all angles.
- Description of the environment where is the equipment and its surrounding: data about the environment (focused on safety), data about others equipments that can influence or be influenced in the maintenance process.

- Description of the users: data about common users of the application, for instance, technicians and engineers.

It is also possible to define strategies and techniques to work upon the data to create more refined data (or information and knowledge) when necessary. Some data may come too "raw" to be used directly in the interface.

The last step of this phase is defining data that will come from an external source. It is important to point which source and what is the technology to communicate with it. For instance: data that will come from sensors in the equipment (watchdog agent, Internet of things), data that will come from E-maintenance systems (via web services) and from wireless sensors networks (WSN).

### B. Creating a behaviour and task model

With all data gathered and structured, now it is time to put it in the form of a task model that can be effectively used to control the flow of the application.

This model is supposed to describe precisely the task execution, step-by-step, using methods like Concur Task Tree (CCT) [16] created for model based design of interfaces.

Not only the task is described, but also every action the user is able to do using the interface. For example, if a help button is available, it has to be described in this model what action it does when clicked.

Besides CCT PDFA (probabilistic deterministic finite automata) will be used to put a weight of probability in each transition and Markov chains to interpret the user history and adapt the interface in real time according to his user model, similar to what is used in [3].

### C. Identifying contexts for situation awareness and team situation awareness

Using the behavior and task model, this phase is to identify contexts of abnormalities that can happen in each part of the task, to increase situation awareness (SA) and team SA.

The idea is to model every possible exception that can be encountered during the task execution, like possible mistakes and risk situations.

If some of the contexts must be interpreted using the data from phase 1, it is possible to define which strategy and data will be used for this interpretation.

Besides that, here is defined the strategy for team SA (if the task can be executed by teams). As stated in [10], in maintenance teams, errors frequently occur because information is not shared or passed between teams.

There are three levels of errors: (1) information unsuccessfully transmitted for the team or team member; (2) information miscomprehended from one person, or comprehended differently between teams; (3) implications of transmitted information for future events miscomprehended across teams or teams' members.

Individuals need not only to comprehend the task they are doing, but also what the team is doing, because this will influence their decision making and performance.

Therefore, when it is a team maintenance, the interface has to reflect for the user at least some indicative of what the team is doing all the time, and it also needs an option for the user to see in more details what each of the team members are doing (and maybe what other teams are doing, if they affect his own work).

### D. Designing user models

With the data gathered, and the tasks and contexts specified, the last phase before defining the interface is defining the user model. There are normally two types of characteristics that go in the user model, application dependent and independent [4].

Application dependent characteristics are: prior experience with computer, knowledge of systems and applications, goals, intentions, expectations, customization parameters, etc.

Application independent characteristics are: neuro-motor skills, preference, capabilities, cognitive level, learning abilities etc.

Four different levels of users exist, and each user is classified in one of these levels. Currently they are denominated: none, basic, advanced and expert. It is important in this phase to determine the strategy to interpret the data and classify each user.

Markov chains will be used during execution to adjust the user level if necessary (hence adjusting the user interface). This will basically function this way: each user level has a probability of completing a task without requesting help, and each task has a time of completion. Based on this, Markov chain can determine the probability of the user being in a certain level, and use this information to change the user level when necessary.

Since each user level has one knowledge and capability of the system, each one has a different interface for executing tasks. One very simplified example could be: level 1 interface (expert) is just textual instruction, level 2 interface (advanced) is visual (photos and images) instructions, level 3 interface (basic) is Augmented Reality instructions and level 4 interface (none) is a video of the task being executed by other user (which can be recorded using the user point of view if a Head-Mounted-Display is being used for AR).

### E. Modeling the user interface

Finally the last step is modeling the user interface. Interfaces will be modeled using components, which were created using patterns of classic AR interfaces components, and 2d components.

Figure 1 demonstrates the entire methodology flow to create an interface: (1)Firstly data is gathered and structured for the complete equipment and surrounding, and the source of the data is also established in case it is dynamic; (2) Then a specific part of the equipment is chosen; (3) The complete maintenance guide for this part is elaborated (beginning of the task model); (4) In this step it is possible to isolate elements of the part, and then for this element mount a task flow, also determining possible exceptions, errors, and Team SA necessities; (5) The creation of the user models is done; (6) All interfaces for all user models are modeled, and also, if wanted, interfaces to view information from the others steps can be specified (or else a standard interface can be used if users wish to view information from other steps). Notice that in Step 6 it is possible to go back and see information about the other steps, therefore the interface is constituted basically by step 4, but the user is able to request and see at anytime information from step 1-3.

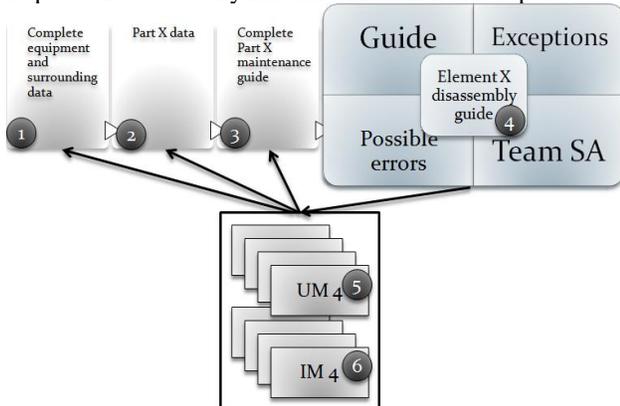

Figure 1: Methodology flow to build an interface

## 5. Discussion and Conclusions

We would like to use this section to discuss some of the work philosophies and decisions.

First of all, why use AR? Because AR is an emerging interface that can be brought to the workplace of nearly every job. When this technology reach its maturity level, contact lenses or high definition optical see through glasses [1] will be available and they will lessen several burdens other interfaces put on users: to use hands to hold things, lack of mobility, the cognitive load of understanding virtual information and applying it in the real world. Also this interface is not limited to projecting 3d objects in the real world, we can and will use 2d interfaces together (as a HUD in users' point of view).

We are trying to achieve an AR interface to improve human work in the maintenance process and to reduce cognitive load through real time data visualization. But carefulness is important, since AR can go from helpful to cumbersome easily if the interface is not well designed.

Continuing the discussion, we believe it is important that the AR application provide safety, following these guidelines [23]:
- Provide the right information in the right time.
- Easy access to information, obtaining during the task, more data about the maintenance of the product or about the failure, and heeds of how to handle the product.
- Improved visualization of warnings in the documentation.
- Update to the AR application when there is an update in the E-maintenance system.

The application should improve efficiency and reduce mistakes with the following guidelines:
- Use up-to-date documentation: documentation can be altered or enhanced to reflect difficulties in interpretation or utilization of certain parts.
- Use check list so that users don't get lost during the process and don't skip steps.
- Reduce the time spent to consult or locate resources and materials.
- Reduce time to report problems: the application has an option to mark that task as problematic. Normal problems could be: incorrect documentation (inconsistent 3D models, tasks missing steps, lacking data, danger not alerted), inaccurate tracking (familiar problem in AR field, also called registration), polluted and clustered interface, insufficient time for system response (may be due to problems in networking, tracking, voice and gesture interpretation).

Applications must also be aware with information about the equipment condition, the environment and the team. If there are risks to humans in the process, we must explore how these risks can be mitigated through better interfaces.

Team work and remote collaboration is also important and an AR interface must be ready to offer solutions for them.

Finally it may be interesting for users to be able to locate pieces or tools in an industrial scenario when necessary. An interface like Pick-by-vision [21] could be used if some form of tracking for the objects is available (IoT and WSN could support this).

Moreover, our objective is that not only maintenance tasks can be facilitated by the AR applications, but also any procedural task dealing with complex industrial equipment. Some of the activities that could be supported, as appointed by Henderson and Feiner [11] are: inspection, testing, servicing, alignment, installation, removal, assembly, repair, overhaul and rebuilding.

Even further, training could be done using the result of this work. Since AR can display virtual information in the real world, trainings could be designed as assemblies' tasks simulated putting virtual problems in real equipments. Also virtual immersive trainings could be done, using the model generated from the methodology to create augmented reality in virtual environments for a virtual training.

We finish discussing what happens after the maintenance is accomplished. With the maintenance finished, the AR application should send data back to the E-maintenance system about the process, so that these data can improve the process the next time. These data could be: (1) Data to notify other teams or team members, and release them to do their job in case their task were being blocked by this one; (2) data about the user (name, level of expertise) or the team (integrants), time spent to fulfill the task, any additional information accessed during the task besides the basic provided (virtual help, information from the equipment, from the environment) and situations (contexts) that occurred during the task.

ACKNOWLEDGMENT

The authors wish to thank CNPq and FAPESP for the support to the INCT-SEC Project, processes 573963/2008-8 and 08/57870-9.